\documentclass[twocolumn,english]{article}
\usepackage[T1]{fontenc}
\usepackage{amssymb}
\usepackage{graphicx}

\makeatletter
\usepackage{a4}
\input{epsfig.sty}
\def\fnum@table{\tablename~{\bf\thetable}}
\def\fnum@figure{\figurename~{\bf\thefigure}}
\def\tablename{\footnotesize{\bf Table}}
\def\figurename{\footnotesize{\bf Figure}}

\def\be{\begin{equation}}
\def\ee{\end{equation}}

\makeatother

\usepackage{babel}
\begin{document}

\title{\textbf{Charmed hadron production in the QGSJET-III model: 
relevance to calculations of the prompt atmospheric neutrino background}}

\author{Sergey Ostapchenko and G\"unter Sigl\\
\textit{\small Universit\"at Hamburg, II Institut f\"ur Theoretische
Physik, 22761 Hamburg, Germany}\\
}

\maketitle
\begin{center}
\textbf{Abstract}
\par\end{center}
The treatment of charmed hadron production in the framework of the QGSJET-III
Monte Carlo  generator   is developed, including both  perturbative
and nonperturbative contributions. For the former, a leading order (LO)
perturbative formalism 
is employed. The so-called $K$-factor designed to take effectively into account higher order
corrections, which controls the magnitude of the LO perturbative contribution, is fixed based
on a comparison with experimental data on $D$ meson production at the Large Hadron Collider. 
For the nonperturbative intrinsic charm,
a phenomenological Regge-based treatment is developed, 
 with the corresponding normalization being chosen based on 
  measurements of the spectra of $\Lambda_c$ baryons 
 at large values of Feynman $x$.
 The model is applied for a calculation of the prompt atmospheric muon
 neutrino flux in the $10-100$ TeV and $1-10$ PeV energy intervals.

\section{Introduction\label{intro.sec}}
Secondary charmed hadrons created in hadronic collisions are much less abundant,
compared to the   ones composed of light (anti)quarks. Nevertheless, an accurate
description of charmed hadron production is of importance for astrophysical studies
based on detection of high energy neutrino fluxes \cite{gai95}. This is because the atmospheric
neutrino background created by interactions of high energy cosmic rays (CRs) with air
is dominated above $\sim 1$ PeV by the so-called prompt component, i.e., by neutrinos
resulting from decays of charmed hadrons produced in such interactions. The suppression
of the ``conventional'' atmospheric neutrino flux resulting from decays of pions
and kaons, at PeV energies, is due to the relatively long life time of such particles.
Consequently, very high energy pions and kaons have small chances to decay in flight,
before interacting with air nuclei. In contrast, short life time of charmed hadrons
allows them to decay and to produce thereby an important background for astrophysical
neutrino detection in the PeV energy range.

The production of charm (anti)quarks within the perturbative quantum chromodynamics (pQCD)
framework is rather well understood \cite{nas89,bee91}. Moreover, experimental measurements
of the production spectra of charmed hadrons, in proton-proton collisions, notably, by the
LHCb collaboration \cite{aai13,aai16,aai17}, allowed one to substantially reduce the uncertainties
of the corresponding calculations, related to the low $x$ behavior of parton distribution
functions (PDFs) of gluons   \cite{gau15,zen15,gau17,zen20}
 and concerning the choice of the factorization and
renormalization scales (e.g., \cite{nel13}). Much less constrained is the potential contribution
of the nonperturbative intrinsic charm (IC) \cite{bhps80,bps81}, for which only vague upper bounds
have been derived (e.g., \cite{hob14,jim15,bed19}).

In this work, a treatment of charmed hadron production in the framework of the QGSJET-III
Monte Carlo (MC) generator \cite{ost24a,ost24b} is developed, including both   perturbative
and nonperturbative contributions. For the former, a leading order (LO) pQCD description
is employed. The so-called $K$-factor designed to take effectively into account higher order
corrections, which controls the magnitude of the LO perturbative contribution, is fixed based
on a comparison with LHCb data on $D$ meson production in proton-proton collisions. 
For the nonperturbative intrinsic charm,
a phenomenological Regge-based treatment is developed, 
 with the corresponding normalization being chosen based on measurements of $\Lambda^+_c$ baryon
 production in $pp$ collisions at the Intersecting Storage Rings (ISR) \cite{bar91}.
A possibility to constrain this normalization, using the upper limit on the prompt 
atmospheric neutrino
background, obtained by the IceCube collaboration \cite{aar16}, is discussed.

The outline of the paper is as follows. In Section \ref{pert.sec}, the treatment of the perturbative
charm production is described. Section \ref{nonpert.sec} is devoted to a phenomenological description
of the intrinsic charm contribution. Selected numerical results regarding charm production
in proton-proton collisions are presented in Section \ref{res.sec},
in comparison to experimental measurements of charmed hadron spectra.  The relevance of the
developed formalism to calculations of the prompt atmospheric neutrino background and
the relative importance of the corresponding perturbative and nonperturbative contributions
are discussed in Section \ref{relev.sec}. Finally, the conclusions are given in Section \ref{out.sec}.

\section{Perturbative contribution to charm production\label{pert.sec}}

The QGSJET-III model  \cite{ost24a,ost24b} treats hadronic   collisions within the 
Reggeon Field Theory   framework \cite{gri68,gri69}, describing multiple rescattering processes 
between interacting hadrons (nuclei) as Pomeron exchanges. The latter thus provide an effective
description for microscopic (generally nonperturbative) parton cascades mediating the collisions.
With increasing energy, such cascades enter the perturbative domain of high parton virtualities $q^2$.
This leads one to the concept of   ``semihard Pomeron'': treating the perturbative parton evolution
at $q^2> Q_0^2$, $Q_0^2$ being some cutoff for pQCD being applicable, by means of the  Dokschitzer-Gribov-Lipatov-Altarelli-Parisi (DGLAP) formalism
\cite{gri72,alt77,dok77}, while retaining the phenomenological Pomeron description for the nonperturbative 
(parts of) parton cascades \cite{dre99,dre01,ost02}. Further, one treats nonlinear interaction effects
corresponding to a rescattering of intermediate partons in such cascades off the projectile or target
hadrons (nuclei) or off each other as Pomeron-Pomeron interactions, based on all order resummation of
the respective (so-called enhanced) Pomeron diagrams \cite{ost06,ost08,ost10}. Considering unitarity
cuts of multi-Pomeron graphs describing hadron-proton (hadron-nucleus, nucleus-nucleus) elastic
scattering and applying the so-called Abramovskii-Gribov-Kancheli (AGK) cutting rules \cite{agk74},
one is able to derive both the interaction cross sections for collisions of interest and partial
 probabilities  for various ``macro-configurations'' of such collisions, i.e., for
 various topologies of cut Pomeron ``nets'', where a cut Pomeron represents an elementary production
 process  \cite{ost08,ost10}. In turn, this allows one to employ MC techniques for sampling
  collision events \cite{ost11}.
  
  Very remarkably, by virtue of the AGK cancellations \cite{agk74}, the outcome of the above-discussed
  procedure for parton jet production is consistent with the collinear factorization of pQCD \cite{col89}
  in the sense that the (here LO) inclusive jet production cross section, above a chosen $Q_0^2$-cutoff,
   is described by the usual pQCD
  factorization ansatz\footnote{Modulo higher twist corrections \cite{ost24a}.} \cite{ost06a,ost16}:
  \begin{eqnarray}
\sigma^{\rm jet}_{ab}(s,Q_0^2)=K_{\rm f}\int \!dx^+dx^-d\hat t
 \; \frac{\pi\,\alpha_{\rm s}^2(\mu_{\rm R}^2)}{\hat s^2}
 \nonumber \\
\times  \sum_{I,J,I',J'} \frac{1}{1+\delta_{I'J'}}\, f_{I/a}(x^+,\mu_{\rm F}^2)\,  f_{J/b}(x^-,\mu_{\rm F}^2) \nonumber \\
\times  \;|M_{IJ\rightarrow I'J'}(\hat s,\hat t)|^2\,\Theta (\mu_{\rm F}^2-Q_0^2), \label{eq:sigjet}
\end{eqnarray}
where $s$ is the center-of-mass (c.m.) energy squared for hadron-proton (nucleon-nucleon) interaction,
$\hat s$ and $\hat t$ are Mandelstam variables for parton-parton scattering, $x^{\pm}$ are  light cone 
(LC) momentum fractions of partons $I$ and $J$, respectively, $\alpha_{\rm s}$ is the running coupling,
$f_{I/a}$ is the PDF of parton $I$ in particle $a$, $M_{IJ\rightarrow I'J'}$ is the LO
matrix element for the Born $IJ\rightarrow I'J'$ scattering, $\mu_{\rm F}$ and  $\mu_{\rm R}$ are the
factorization and renormalization scales ($\mu_{\rm F}=\mu_{\rm R}=p_{\perp}/2$ is used, with $p_{\perp}$
being the transverse momentum of  partons $I'$ and $J'$, produced in the hard process),
and the factor $K_{\rm f}=1.5$ is designed to take effectively into account higher order pQCD corrections.
It is noteworthy that in the above-described model treatment, the PDFs $f_{I/a}$ comprise absorptive
corrections due to intermediate parton rescattering off their parent hadrons (nuclei) \cite{ost24a,ost06a}.
 
 The fragmentation of final partons into hadrons involves $s$-channel parton cascades (so-called
 final state radiation) and a hadronization of strings of color field stretched between those
 partons or/and constituent partons of the interacting hadrons (nuclei) \cite{ost24a,ost24b}.
 
 While secondary particle production in the QGSJET-III MC generator was restricted to hadrons
 composed of light ($u$, $d$, and $s$) (anti)quarks, it can be trivially generalized to include
 perturbative charm production by adding $gg\rightarrow c\bar c$ and  $q\bar q\rightarrow c\bar c$
 subprocesses to the right-hand-side 
 of Eq.\ (\ref{eq:sigjet}), with the corresponding LO
 matrix elements squared (e.g., \cite{ell96})
  \begin{eqnarray}
|M_{gg\rightarrow c\bar c}(\hat s,\hat t)|^2=\left(\frac{1}{6\tau_1\tau_2} - \frac{3}{8}\right)   \nonumber \\
\times \left(\tau_1^2+\tau_2^2+\rho -\frac{\rho^2}{4\tau_1\tau_2}\right) , \label{eq:gg-cc} &&\\
|M_{q\bar q\rightarrow c\bar c}(\hat s,\hat t)|^2=\frac{4}{9}\left(\tau_1^2+\tau_2^2+\rho/2\right) . \label{eq:qq-cc} &&
\end{eqnarray}
Here $\tau_1= (m_c^2-\hat t)/\hat s$, $\tau_2= (\hat s +\hat t-m_c^2)/\hat s$,
 $\rho =4m_c^2/\hat s$, and 
$m_c$ is the $c$-quark mass (we use $m_c=1.3$ GeV).

At this step, we choose the factorization and renormalization scales for $c$-quark production as
$\mu^c_{\rm F}=\mu^c_{\rm R}=2m_{c,\perp}$, where $m_{c,\perp}=\sqrt{m_c^2+p_{\perp}^2}$, i.e.,
differently compared to the ones for gluon and light quark production. Correspondingly, we use
here a different $K$-factor, $K_{\rm f}^c=3$, whose value has been fixed based on a 
comparison with LHCb
data on $D^0+\bar D^0$ production  (see Section \ref{res.sec}). Generally,
in a LO approach,  changes of $\mu_{\rm F}$ and $\mu_{\rm R}$ can be more or less absorbed
into a suitable redefinition of the $K$-factor. Additionally, for our choice of  
$\mu^c_{\rm F}$ and $\mu^c_{\rm R}$,  the treatment of charm production is independent of the
technical parameter of the model, the $Q_0^2$-cutoff ($Q_0^2=2$ GeV$^2$ is used).

Regarding   hadron production, including charmed secondaries requires only minor changes
in the corresponding MC procedure. After determining the ``macro-configuration'' of a collision, 
we proceed as usual with energy-momentum sharing between elementary production processes and
with a simulation of $t$-channel parton cascades (so-called initial state radiation) for
semihard scattering processes, using a forward evolution algorithm \cite{ost24a}. Restricting 
ourselves with three active flavours, the only change is the possibility to produce a 
$c\bar c$-pair in the hardest scattering process.
In  the three flavour evolution scheme employed,   produced charm (anti)quarks do not undergo
final state cascading. The procedure completes with fragmentation of strings of color field
stretched between final partons. 

The string fragmentation is modeled as an iterative emission of secondary hadrons from
string ends, with the LC momentum fraction $x$ of a hadron (LC$^+$ for the 
projectile side end and  LC$^-$ for the  target side end),
 being sampled according to the distribution 
 \begin{equation}
f_{qq'\rightarrow h}(x) \propto x^{1-\alpha_q-\alpha_{q'}}(1-x)^{\Lambda-\alpha_{q'}} , \label{eq:fragm}
\end{equation}
where $q$ is the string end parton, $q'$ is the parton created from the vacuum to form
hadron $h$, and the parameters $\alpha_q$ are expressed via intercepts of the relevant
Regge trajectories \cite{ost24b}. For charm (anti)quark as the string end, 
 $\alpha_q=\alpha_c=\alpha_{\psi}(0)\simeq -2$ \cite{kai03} and we consider the 
 production of $\Lambda_c$ (anti)baryons ($\alpha_{q'}=\alpha_{ud}=2\alpha_N(0)-\alpha_{\mathbb R}$)
 and $D$-mesons ($\alpha_{q'}=\alpha_{u}=\alpha_{d}=\alpha_{\mathbb R}$),
  with $\alpha_N(0)=0$ and $\alpha_{\mathbb R}=0.5$ \cite{kai86,kai87}.
  Contributions of the production and decays of heavier charmed hadrons are assumed
  to be effectively taken into account in the $\Lambda_c$ (anti)baryon and  $D$ meson
  yields, via the duality principle.

\section{Treatment of nonperturbative intrinsic charm \label{nonpert.sec}}

In addition to the perturbative mechanism for charm production, discussed in Section \ref{pert.sec},
an important contribution to spectra of secondary charmed hadrons may be related to the  intrinsic charm
content of hadrons \cite{bhps80,bps81}.  In that regard, it is important to keep in mind
that  IC (anti)quarks are present in hadrons in a virtual state, being a part of multiparton
fluctuations, constituent parton Fock states, ``prepared'' well in advance of a hadron-proton
(hadron-nucleus, nucleus-nucleus) collision  \cite{bhps80}. It is therefore the interaction process
which can ``awake'' the  intrinsic charm, putting the $c$ and $\bar c$ on shell. In our treatment,
we follow the approach of \cite{bro89,bro92}, assuming that an inelastic interaction of {\em any} parton
 constituent of
incident hadron is sufficient to destroy the coherence of the initial parton fluctuation
and, in case that fluctuation contains a $c\bar c$ state, 
 ``to free the intrinsic charm from its virtual state'' \cite{bro92}.
Thus, the IC   content of a hadron can be characterized
 by a single parameter,  namely, by the relative
weight of constituent Fock states containing such  $c\bar c$ pairs \cite{ogs23}.
 In our approach, this is
slightly modified since we consider a Good-Walker (GW) decomposition of hadron wave 
functions \cite{goo60}, with different GW states being characterized by different spatial sizes
and different (integrated) parton densities \cite{ost24a}. It is natural to expect that in
more compact Fock states, the characteristic virtuality of constituent partons is higher,
$q^2_{\rm const} \propto 1/R^2_{h(i)}$, $R_{h(i)}$ being the transverse size of state $|i\rangle$
of hadron $h$ \cite{fra08}. Therefore, we assume that a compact $c\bar c$ pair is more likely
to emerge in a smaller size state, defining the probability for IC to be present in state 
 $|i\rangle$ of hadron $h$ as
 \begin{equation}
w^{\rm IC}_{h(i)}=w^{\rm IC}_{0}\, R^2_{p(1)}/R^2_{h(i)}, \label{eq:w_ic}
\end{equation}
where $R_{p(1)}$ is the size of the largest GW state considered
 of the proton and $w^{\rm IC}_{0}$ is the
IC content of that state. Based on a comparison with ISR data on $\Lambda_c^+$ baryon
production, we choose  $w^{\rm IC}_{0}=3\cdot 10^{-4}$ (see Section \ref{res.sec}).
It is noteworthy that, since the relative role of small size GW states increases with 
energy \cite{ost24a}, the above-discussed assumptions may lead to a slightly faster energy rise
of the IC contribution to charmed hadron production, compared to the energy dependence
of the inelastic cross section for a reaction.

Regarding the LC momentum fractions $x$ possessed by IC (anti)quarks, we rely on a
Regge-based approach: assuming that the corresponding distribution follows the
Regge asymptotic, $\propto x^{-\alpha_{\psi}(0)}$, in the low $x$ limit  \cite{kai03}.
As demonstrated in \cite{ogs23}, the corresponding momentum distribution is 
not too different from the one of the original  
 Brodsky-Hoyer-Peterson-Sakai model  \cite{bhps80,bps81}.

The next crucial point concerns the hadronization procedure for IC  (anti)quarks.
For an incident proton, we assume that the $c$ quark is combined with the valence $ud$
diquark, forming a  $\Lambda_c^+$ baryon, which is supported by 
  measurements of  $\Lambda_c^+$ spectra  at large values of Feynman $x$ \cite{bar91,cha87,gar02}.
  On the other hand, the remaining valence quark and the $\bar c$ are positioned at the ends
  of strings of color field, resulting from the inelastic scattering process.
  For incident pions and kaons, we assume a formation of 
  a $D$ meson by the $\bar c$ ($c$) quark
  and a light valence (anti)quark, while the remaining valence antiquark (quark) and the
   $c$ ($\bar c$) work as string ends. Thus, for hadron $h$ with active  intrinsic charm,
    undergoing $n$  inelastic   rescatterings, the LC momentum partition between
     $2n$ string end partons and the ``leading'' charmed hadron $h_c$ is described by the
     distribution
 \begin{eqnarray}
 f_h^{\rm LC}(x_1,...,x_{2n},x_{h_c})\propto \left[\prod_{i=1}^{2n-2}
 x_i^{-\alpha_{\rm sea}}  \right] \nonumber \\
 \times \; x_{2n-1}^{-\alpha_{q_{\rm v}}}\, x_{2n}^{-\alpha_{\psi}(0)}\,
 x_{h_c}^{-\alpha_{h_c}}\, \delta \!\left(1-x_{h_c}-\sum_{j=1}^{2n}x_j\right) .
 \label{ems.eq}
 \end{eqnarray}
Here the exponent $\alpha_{\rm sea}$ is used for sea (anti)quarks
while the light valence
 (anti)quark and the IC $c$ ($\bar c$) follow the Regge  behavior: 
 $\propto  x_{2n-1}^{-\alpha_{q_{\rm v}}}$ ($\alpha_{u}=\alpha_{d}=\alpha_{\mathbb R}$,
 $\alpha_{s}=\alpha_{\phi}(0)=0$) and  $\propto x_{2n}^{-\alpha_{\psi}(0)}$, respectively.
 For $h=p$, $h_c=\Lambda_c^+$ and
   \begin{equation}
\alpha_{\Lambda_c}=\alpha_{ud}+\alpha_{\psi}(0)-1=2\alpha_N(0)-\alpha_{\mathbb R}+\alpha_{\psi}(0)-1   , \label{eq:alpha-lamc}
\end{equation}
  while for $h$ being a pion or kaon, a leading $D$ meson is produced, with
  $\alpha_{D}=\alpha_{\mathbb R}+\alpha_{\psi}(0)-1$.
    The fragmentation of all final strings, including
   ones with $c$ or $\bar c$ at their ends, is treated as described in Section \ref{pert.sec}.
 The parameters involved have
  been specified above or listed in \cite{ost24b}.

\section{Selected numerical results for charm production\label{res.sec}}
\begin{figure}[htb]
\centering
\includegraphics[height=6.cm,width=0.48\textwidth]{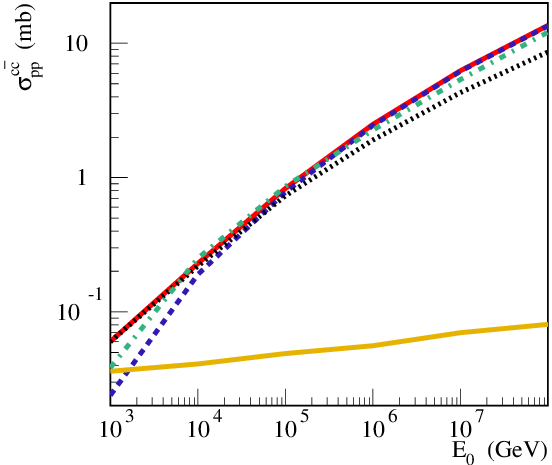}
\caption{Laboratory  energy dependence of the cross section
for  charm production in $pp$ collisions,
$\sigma_{pp}^{c\bar c}$,  calculated using the default model settings
 (red solid line); partial contributions of the perturbative 
 and nonperturbative  mechanisms to $\sigma_{pp}^{c\bar c}$ 
  are shown by blue dashed and yellow solid lines, respectively.
 $\sigma_{pp}^{c\bar c}$ calculated  using $\mu^c_{\rm F}=\mu^c_{\rm R}=1.2 m_{c,\perp}$
   and  $K_{\rm f}^c=2$ in the pQCD treatment is plotted by black dotted line
   while the results of Ref.\ \cite{bha16} are shown by green dash-dotted line.}
\label{fig:sig-cc}       
\end{figure}

 In Fig.\ \ref{fig:sig-cc}, we plot the   energy dependence 
 of the cross section for $c\bar c$ pair production in proton-proton collisions,
 $\sigma_{pp}^{c\bar c}$, calculated using the above-described formalism, showing
 also partial contributions of the perturbative and nonperturbative  mechanisms\footnote{IC 
 contributions both from the projectile and target protons are accounted for.}.
 It is easy to see that the latter depends rather weakly on   energy, dominating the
 charm production  at relatively low energies only, while constituting a small fraction
  of $\sigma_{pp}^{c\bar c}$ in the very high energy limit. This is not surprising
 since the  intrinsic charm is a property of a hadron wave function, hence, the cross
 section for IC activation constitutes an approximately constant fraction of the total
 inelastic cross section for a reaction (see, e.g., the corresponding discussion in \cite{ogs23}).
 For comparison, we also plot in  Fig.\ \ref{fig:sig-cc}  $\sigma_{pp}^{c\bar c}$ calculated
 in \cite{bha16}, using a next-to-leading order (NLO) DGLAP formalism, with the factorization
 and renormalization scales, $\mu^c_{\rm F}=2.1 m_{c,\perp}$ and $\mu^c_{\rm R}=1.6 m_{c,\perp}$,
  being adjusted to fit LHCb data on charmed hadron production \cite{nel13}. 
  At sufficiently high energies,
  our results for the perturbative contribution to 
   $\sigma_{pp}^{c\bar c}$, shown by the dashed line in the Figure,
   are very close to the ones of Ref.\  \cite{bha16},
   which illustrates the importance of those LHCb measurements for 
   constraining both LO and NLO pQCD predictions. 
   
   It is worth repeating that for our LO treatment, 
   we have a certain degeneracy 
   between the
   choice of the $K$-factor,  $K_{\rm f}^c$, and of the factorization and renormalization scales,
  $\mu_{\rm F}^c$ and $\mu_{\rm R}^c$, such that   changes of the latter two   
   can be more or less absorbed into a suitable redefinition of the former. 
   This is illustrated in Fig.\  \ref{fig:sig-cc}, where we plot by the dotted line the charm 
   production cross section obtained using $\mu^c_{\rm F}=\mu^c_{\rm R}=1.2 m_{c,\perp}$
   and  $K_{\rm f}^c=2$ for the perturbative contribution.
    In such a case, however, we have a somewhat slower energy rise of   $\sigma_{pp}^{c\bar c}$: 
    since for a lower $\mu^c_{\rm F}$, the  gluon PDF is  probed
   at a lower virtuality scale, where it is characterized by a less steep low $x$ rise.
   
\begin{figure*}[t]
\centering
\includegraphics[height=6.cm,width=0.9\textwidth]{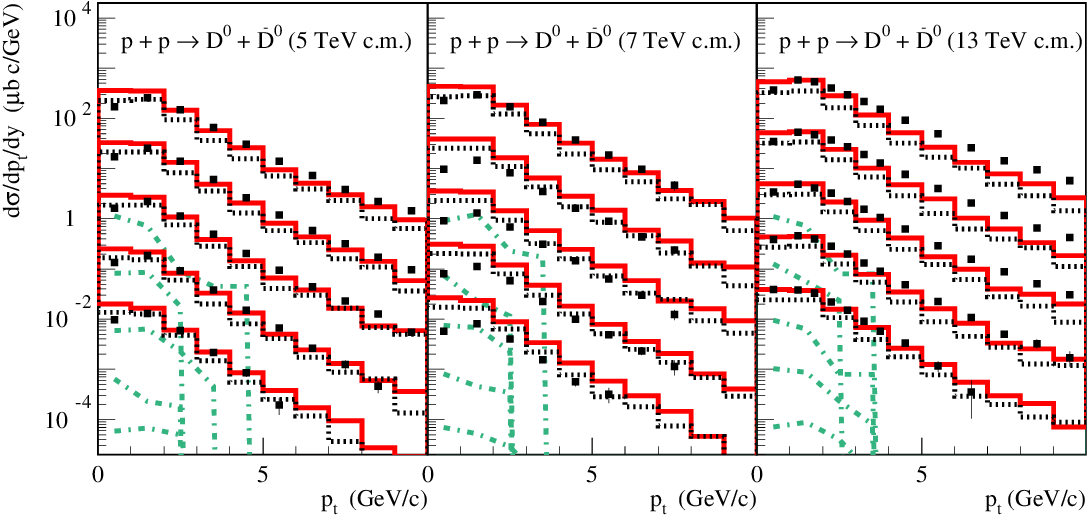}
\caption{Transverse momentum dependence of  $D^0+\bar D^0$ production cross sections,
for $pp$ collisions at  $\sqrt{s}=5$, 7, and 13 TeV -- left, middle, and right panels,
respectively, for different rapidity intervals [from top to bottom: $2<y<2.5$
 ($\times 10^0$),  $2.5<y<3$ ($\times 10^{-1}$),  $3<y<3.5$ ($\times 10^{-2}$),  $3.5<y<4$
  ($\times 10^{-3}$),  $4<y<4.5$ ($\times 10^{-4}$)]. Solid histograms 
  correspond to   calculations using the default model parameters, 
with the partial contributions of IC being plotted as green dash-dotted  lines,
 while  dotted  histograms are obtained using
   $\mu^c_{\rm F}=\mu^c_{\rm R}=1.2 m_{c,\perp}$
   and  $K_{\rm f}^c=2$ in the pQCD treatment.
    LHCb data  \cite{aai13,aai16,aai17} are shown by points.}
\label{fig:lhcb-pt}       
\end{figure*}

   Regarding a comparison with accelerator data, we start in Fig.\  
     \ref{fig:lhcb-pt} 
 with confronting our calculations  of transverse
      momentum spectra, for a number of rapidity intervals, of  $D^0+\bar D^0$ mesons produced in $pp$ collisions
       at $\sqrt{s}=5$, 7,  and 13 TeV to LHCb measurements\footnote{The measured $D^0+\bar D^0$ 
       yields comprise contributions of $D^*$ decays.}   \cite{aai13,aai16,aai17}.
  While we have an acceptable overall agreement with the data, the calculated cross sections
  fall down somewhat faster with increasing  $p_{\perp}$, than observed by the experiment.
  We also plot as dotted histograms in  Fig.\       \ref{fig:lhcb-pt} the results of our calculations using 
      $\mu^c_{\rm F}=\mu^c_{\rm R}=1.2 m_{c,\perp}$  and  $K_{\rm f}^c=2$ in the pQCD
      treatment. Clearly, they
      are rather similar to our default case, except that the slower energy rise of the
      perturbative charm      production for these settings improves somewhat the agreement with the data
      at  $\sqrt{s}=7$  TeV, while leading at the same time to a sizable underestimation of the
        $D^0+\bar D^0$ yield   at $\sqrt{s}=13$  TeV.
 
\begin{figure}[htb]
\centering
\includegraphics[height=6.cm,width=0.48\textwidth]{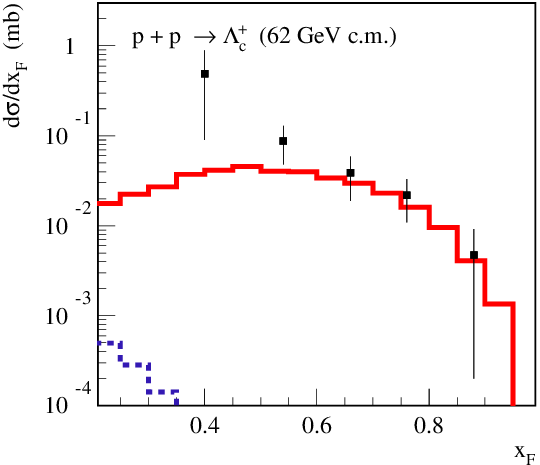}
\caption{Calculated Feynman $x$ dependence  of the production cross section for  $\Lambda_c^+$,
for $pp$ collisions at  $\sqrt{s}=62$ GeV, and the  partial contribution of the perturbative 
  charm production -- red solid and blue dashed histograms, respectively, compared to
  ISR data  \cite{bar91} (points).}
\label{fig:isr-x}       
\end{figure}

\begin{figure}[htb]
\centering
\includegraphics[height=6.cm,width=0.48\textwidth]{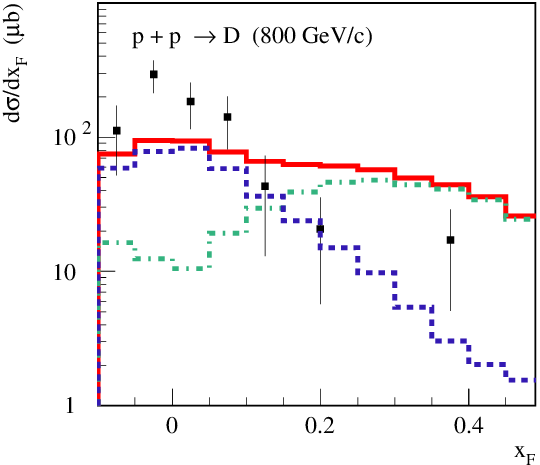}
\caption{Calculated Feynman $x$ dependence  of the production cross section for  $D$ mesons,
for $pp$ collisions at 800 GeV/c, and   partial contributions of the perturbative and nonperturbative
   production mechanisms -- red solid, blue dashed, and green dash-dotted histograms, respectively, compared to
 the  data of the LEBC-MPS collaboration \cite{amm88} (points).}
\label{fig:d0-800}       
\end{figure}

 Concerning the intrinsic charm, its contribution to the results plotted in 
   Fig.\   \ref{fig:lhcb-pt}, shown by the dash-dotted lines in the Figure, proves to be miserable.
    Therefore, the IC content of the proton can not be reliably constrained by
   those LHCb data. In that regard, our main experimental constraint 
   comes from measurements of  $\Lambda_c^+$ baryon production at large values of
   Feynman $x$ in $pp$ collisions at ISR \cite{bar91}. Our calculations are compared to the 
   data on    Feynman $x$  distribution of  $\Lambda_c^+$ in 
   Fig.\  \ref{fig:isr-x}. 
 Clearly, the calculated  $\Lambda_c^+$ 
   yield comes predominantly from the  intrinsic charm, with the perturbative
   contribution shown by the dashed line  in the Figure, being negligibly small in the
   kinematic range studied.
    While the error bars for the
   experimental data are rather large, an additional constraint on the IC   content
    is provided by  measurements of  $D$ meson production 
   in $pp$ collisions at 800 GeV/c by the LEBC-MPS collaboration  \cite{amm88}, our calculations being
   compared to the data in Fig.\   \ref{fig:d0-800}.
   Here it is easy to see that our normalization
   for  the IC   content  of the proton likely corresponds to the upper bound:
    the calculated  $x_{\rm F}$ dependence of    $D$ meson  yield is systematically flatter than the  observed one.
 
\section{Relevance to calculations of the prompt atmospheric neutrino flux\label{relev.sec}}

Regarding the relevance of charmed hadron production to calculations of the atmospheric neutrino background,
of importance is not the total $c\bar c$ pair production cross section plotted in Fig.\ \ref{fig:sig-cc} but rather
the yield of charmed hadrons at large values of Feynman $x$, $x_{\rm F}\gtrsim 0.1$, which is a consequence
of the steeply falling down primary CR spectrum. In particular, in the energy range where the primary
proton flux can be approximated by a power law,
\begin{equation}
 I_p(E_0)\simeq I_p(E_{\rm ref})\, (E_0/E_{\rm ref})^{-\gamma_p}\,, \label{eq:p-flux}
\end{equation}
with $E_{\rm ref}$ being some reference energy and $\gamma_p$ -- the spectral slope,  the proton contribution to the
prompt neutrino background is proportional to
 spectrum-weighted moment of the distribution of charm (anti)quarks produced
in proton-air collisions \cite{ogs23}, the latter being defined as
\begin{equation}
Z_{p{\rm -air}}^c(E,\gamma_p)=\int \!dx\,x^{\gamma_p-1}\,
\frac{dn_{p{\rm -air}}^c(E/x,x)}{dx}\,.
\label{eq:c-gener}
\end{equation}
Here $dn_{p{\rm -air}}^c/dx$ is the $c+\bar c$ distribution,
with respect to the energy fraction $x=E/E_0$ taken  from the parent proton.
 
\begin{figure}[htb]
\centering
\includegraphics[height=6cm,width=0.48\textwidth]{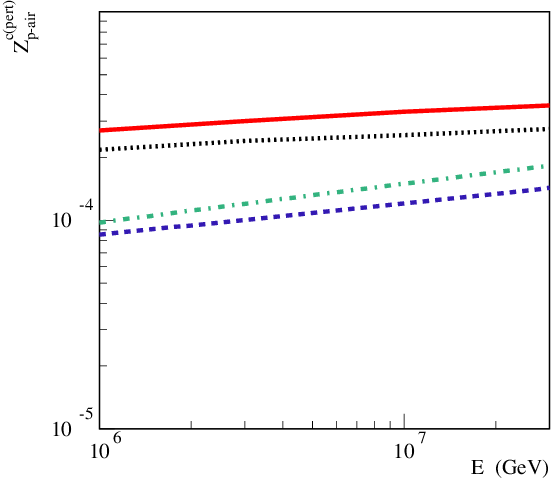}
\caption{Energy dependence of  $Z_{p{\rm -air}}^{c{\rm (pert)}}$, for  $\gamma_p=3$,
 calculated with the default model settings or using 
 $\mu^c_{\rm F}=\mu^c_{\rm R}=1.2 m_{c,\perp}$,  $K_{\rm f}^c=2$
 -- red solid and black dotted  lines, respectively.
The results of NLO calculations based on gluon PDFs from
CT14nlo\_NF3 \cite{ct14} and  NNPDF31\_nlo\_pch\_as\_0118\_nf\_3 
 \cite{bal11}  PDF sets
are plotted by, correspondingly, blue dashed and green dash-dotted lines.}
\label{fig:Zc-pert}       
\end{figure}

 The energy dependence of the perturbative contribution to 
  $Z_{p{\rm -air}}^c(E,\gamma_p)$,  $Z_{p{\rm -air}}^{c{\rm (pert)}}(E,\gamma_p)$,
  for $\gamma_p=3$, calculated
 both for the default model settings and with  
  $\mu^c_{\rm F}=\mu^c_{\rm R}=1.2 m_{c,\perp}$,   $K_{\rm f}^c=2$, 
  is plotted in Fig.\    \ref{fig:Zc-pert}. For comparison,  the
corresponding results of  NLO pQCD calculations using different gluon PDFs,
described in some detail in \cite{ogs23},
for    $\mu^c_{\rm F}=\mu^c_{\rm R}= m_{c,\perp}$,
are shown.
The  differences between the presented results for  $Z_{p{\rm -air}}^{c{\rm (pert)}}$
are of the same order as the theoretical uncertainties of NLO calculations,
estimated by varying the factorization and renormalization scales (see, e.g., Fig.\ 6
in  \cite{ogs23}), and may
give one some feeling about the uncertainties of the predictions
for  the  prompt atmospheric neutrino flux, related to the perturbative
treatment of charm production.
 
 While our full scale calculation of the  atmospheric neutrino background will be presented
 elsewhere, here we can provide rough estimations for the prompt contribution to the 
 atmospheric muon neutrino (plus antineutrino) flux, $I_{\nu_{\mu}{\rm (prompt)}}$,
  both well below the energy of the so-called  ``knee'' of the primary CR spectrum,
 $E_{\rm knee}\simeq 3-4$ PeV \cite{kul59}, 
 and above  the  ``knee'', i.e., in the energy intervals where the relevant primary CR
 spectra can be  approximated by a power law behavior.
 
 Starting with the  power law proton flux, Eq.\ (\ref{eq:p-flux}), and proceeding like in
 \cite{ogs23} but factoring out only charmed hadron decay moments, we get for the primary
 proton contribution to $I_{\nu_{\mu}{\rm (prompt)}}$:
 \begin{eqnarray}
I_{\nu_{\mu}{\rm (prompt)}}^{(p)}(E)\simeq I_p(E_{\rm knee})\,
\frac{(E/E_{\rm knee})^{-\gamma_p}} 
{1-Z_{p{\rm -air}}^N(E,\gamma_p)} 
&&\nonumber \\
\times\; \sum_{h_c} 
Z_{p{\rm -air}}^{h_c}(E,\gamma_p)\;
Z^{\rm dec}_{h_c\rightarrow \nu_{\mu}}(\gamma_p) \,,&&
\label{eq:Inu-prompt-p}
\end{eqnarray}
where we use $E_{\rm ref}=E_{\rm knee}$ and  $Z_{p{\rm -air}}^N$ is the
spectrum-weighted moment for nucleon ``regeneration'':
\begin{equation}
Z_{p{\rm -air}}^N(E,\gamma_p)=\int \!dx\,x^{\gamma_p-1}\,
\frac{dn_{p{\rm -air}}^N(E/x,x)}{dx}\,,
\label{eq:p-regener}
\end{equation}
with $dn_{p{\rm -air}}^N/dx$ being the  distribution of secondary nucleons
 in proton-air collisions.

In turn,   $Z_{p{\rm -air}}^{h_c}$ is the  spectrum-weighted moment for charmed hadron $h_c$
production in  proton-air collisions 
while $Z^{\rm dec}_{h_c\rightarrow \nu_{\mu}}$ is the moment for  $h_c$ decay into $\nu_{\mu}$:
\begin{eqnarray}
Z_{p{\rm -air}}^{h_c}(E,\gamma_p)=\int \!dx\,x^{\gamma_p-1}\,
\frac{dn_{p{\rm -air}}^{h_c}(E/x,x)}{dx} &&
\label{eq:hc-gener} \\
Z^{\rm dec}_{h_c\rightarrow \nu_{\mu}}(\gamma_p)=
 \int  \!dx\,x^{\gamma_p-1}\,
f^{\rm dec}_{h_c\rightarrow \nu_{\mu}}(x)\,. &&
\label{eq:z-dec}
\end{eqnarray}
Here  $f^{\rm dec}_{h_c\rightarrow \nu_{\mu}}(x)$ is the distribution for
 the energy fraction  of   $h_c$, taken by the neutrino, multiplied by the
 corresponding branching ratio.

    \begin{figure*}[t]
\centering
\includegraphics[height=6cm,width=\textwidth]{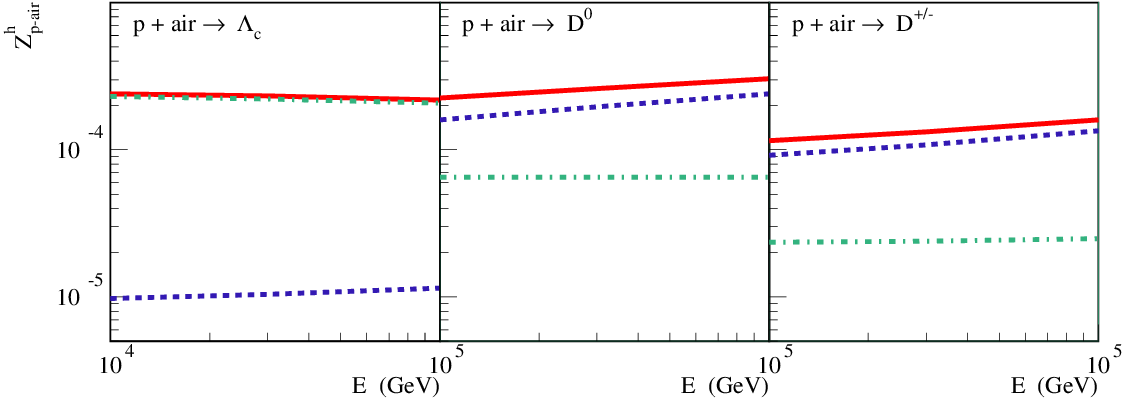}
\caption{Calculated energy dependence of  $Z_{p{\rm -air}}^{h_c}$, for  $\gamma_p=2.7$,
for $h_c=\Lambda_c+\bar \Lambda_c$ (left panel),  $h_c=D^0+\bar D^0$ (middle panel), 
and   $h_c=D^{\pm}$ (right panel) -- red solid lines. Partial contributions
 of the  perturbative and nonperturbative mechanisms are plotted 
 as blue dashed and green dash-dotted lines, respectively.}
\label{fig:zmom-2.7}       
\end{figure*}%
It is noteworthy that Eq.\ (\ref{eq:Inu-prompt-p}) is applicable both for the perturbative and IC
contributions to the neutrino flux since we have not factorized out the fragmentation functions
of charmed (anti)quarks into hadrons. This is quite important because we consider different
hadronization mechanisms in the two cases.

In Fig.\ \ref{fig:zmom-2.7}, we plot, for $\gamma_p=2.7$,
      \begin{figure*}[t]
\centering
\includegraphics[height=6cm,width=\textwidth]{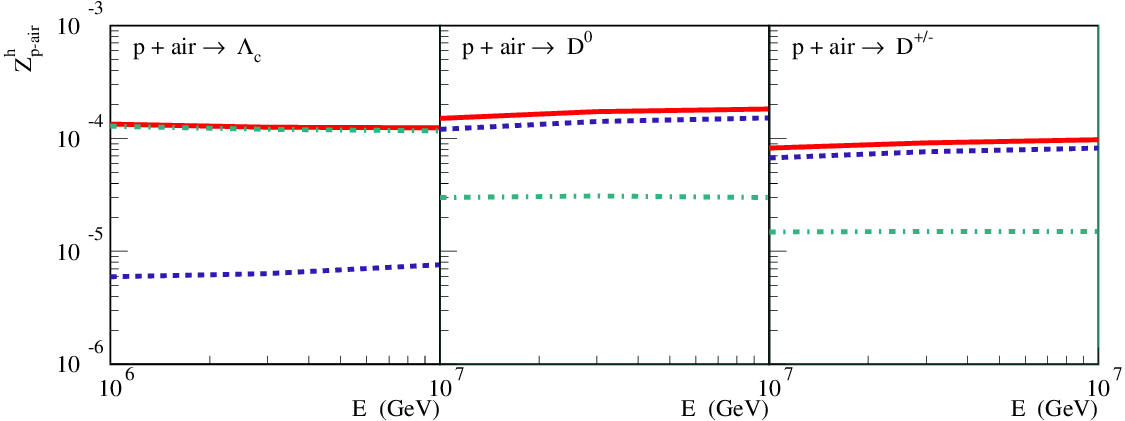}
\caption{Same as in  Fig.\ \ref{fig:zmom-2.7}, for  $\gamma_p=3.2$.}
\label{fig:zmom-3.2}       
\end{figure*}%
 the energy dependence of the moments 
 $Z_{p{\rm -air}}^{h_c}$, between 10 and 100 TeV,
for $h_c=\Lambda_c$ ($+\bar \Lambda_c$), $D^0$ ($+\bar D^0$), and  $D^{\pm}$,
  showing also  the partial contributions of the 
 perturbative and nonperturbative mechanisms.
 The same energy dependence at higher energies, between 1 and 10 PeV, is plotted
 in Fig.\   \ref{fig:zmom-3.2}, for $\gamma_p=3.2$.
  The two considered values of $\gamma_p$ are characteristic for
 the slopes of the primary CR proton spectra below and above the ``knee'', respectively
  \cite{ant05,ape11,ape13b,ape13a,aar19}.
 While the perturbative mechanism provides the dominant contribution to the calculated
 moments for $D$ meson production, for the chosen model parameters, an important
 correction, $\sim 20-30$\%, depending on the energy, comes from
 the IC content of the proton. On the other hand, regarding $\Lambda_c$ production,
 the corresponding moment  $Z_{p{\rm -air}}^{\Lambda_c}$ is almost entirely defined
 by intrinsic charm, with the perturbative mechanism contributing less than 10\%.
  In contrast to  $\sigma_{pp}^{c\bar c}$
 plotted in  Fig.\ \ref{fig:sig-cc}, where intrinsic charm provides a minor correction
 at the high energies of interest, its relative contribution to the CR spectrum-weighted 
 moments, Eq.\ (\ref{eq:hc-gener}), is enhanced due to the much harder energy spectra
 of secondary charmed hadrons, compared to the ones stemming from the
  perturbative mechanism. Further, as noticed
 previously in  \cite{ogs23},  we have a pretty flat  energy dependence 
 for the IC contributions to these CR spectrum-weighted moments,
 while observing a pronounced 
energy rise of the moments corresponding to the perturbative generation of charm,  driven
by the low $x$ increase of the gluon PDF of the proton. 

Regarding the contributions of primary CR nuclei, those can be well described using the so-called
superposition model, reducing the problem to calculations of the CR
spectrum-weighted moments $Z_{p{\rm -air}}^{h_c}(E,\gamma_A)$ and
 $Z^{\rm dec}_{h_c\rightarrow \nu_{\mu}}(\gamma_A)$,
 using the corresponding slopes $\gamma_A$ for a primary nuclear mass group of interest
  (see, e.g., \cite{ogs23}). Thus, approximating the corresponding partial primary fluxes
  by a power law behavior, either below or above  the respective spectral breaks $E_{\rm knee}^{(A_i)}$,
 \begin{equation}
 I_{A_i}(E_0)\simeq I_{A_i}(E_{\rm knee}^{(A_i)})\:
  (E_0/E_{\rm knee}^{(A_i)})^{-\gamma_{A_i}}\,,
 \label{eq:A-flux}
\end{equation}
$\gamma_{A_i}$ being the corresponding spectral slopes,
  for the total  prompt contribution to the 
 atmospheric muon neutrino flux, we have \cite{ogs23}
\begin{eqnarray}
I_{\nu_{\mu}{\rm (prompt)}}(E)\simeq  \sum_i
\frac{A_i^{2-\gamma_{A_i}}\;I_{A_i}(E_{\rm knee}^{(A_i)})}
{1-Z_{p{\rm -air}}^N(E,\gamma_{A_i})}
&&\nonumber \\
\times \!\!\left[\frac{E}{E_{\rm knee}^{(A_i)}}\right]^{-\gamma_{A_i}}
\!\!  \sum_{h_c} 
Z_{p{\rm -air}}^{h_c}(E,\gamma_{A_i})\,
Z^{\rm dec}_{h_c\rightarrow \nu_{\mu}}(\gamma_{A_i}).&&
\label{eq:Inu-prompt-tot}
\end{eqnarray}
In practice, contributions of primary nuclei heavier than helium can be neglected at 10\%
accuracy level.

Based on  the data of the  KASCADE-Grande \cite{ape13b} and Ice-Cube \cite{aar19} experiments, we  parametrize the primary  fluxes of protons  and 
helium above the corresponding ``knees'' as (see Fig.\ \ref{fig:cr-spec}):
    \begin{figure}[htb]
\centering
\includegraphics[height=4.5cm,width=0.48\textwidth]{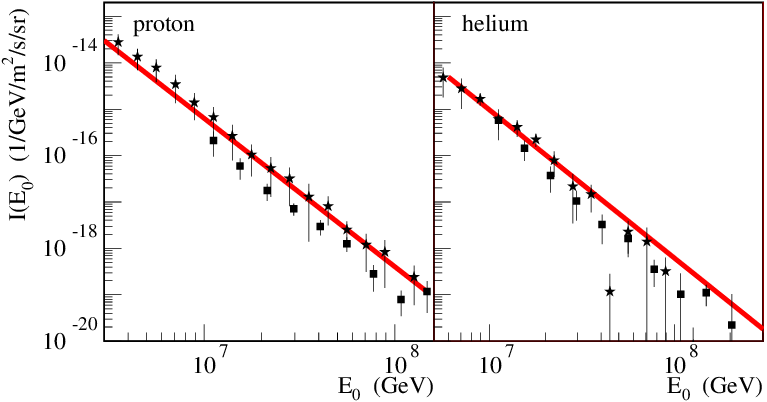}
\caption{Parametrizations of the primary fluxes of protons (left) and 
helium (right) -- red solid lines,
 compared to experimental data of KASCADE-Grande \cite{ape13b}
(squares) and Ice-Cube \cite{aar19} (stars).}
\label{fig:cr-spec}       
\end{figure}%
\begin{eqnarray}
 I_p(E_0)=I_p(E^p_{\rm knee})\, (E_0/E^p_{\rm knee})^{-3.2}\nonumber \\
 \times \,\Theta (E_0-E^p_{\rm knee})  \label{eq:p-flux-high}\\
 I_{\rm He}(E_0)=I_p(E^{\rm He}_{\rm knee})\, (E_0/E^{\rm He}_{\rm knee})^{-3.2}\nonumber \\
 \times \,
 \Theta (E_0-E^{\rm He}_{\rm knee})\, ,  \label{eq:he-flux-high}
\end{eqnarray}
with $E^p_{\rm knee}=3$ PeV, $E^{\rm He}_{\rm knee}=2E^p_{\rm knee}$,
 $I_p(E^p_{\rm knee})=3\times 10^{-14}$ GeV$^{-1}$m$^{-2}$s$^{-1}$~sr$^{-1}$, and
 $E^{\rm He}_{\rm knee}= 5\times 10^{-15}$ GeV$^{-1}$m$^{-2}$s$^{-1}$~sr$^{-1}$.
 Below the ``knees'', we extend these primary spectra, using the slope
 equal $-2.7$, 
which agrees with KASCADE data \cite{ant05,ape13b}, 
within the uncertainties of the experimental analysis.

\begin{figure}[htb]
\centering
\includegraphics[height=5cm,width=0.48\textwidth]{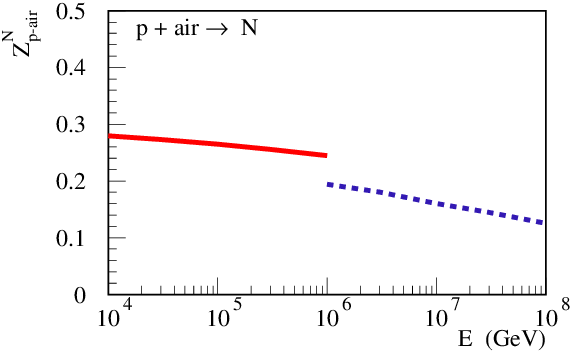}
\caption{Calculated energy dependence of  the nucleon regeneration moment
$Z_{p{\rm -air}}^N$, for $\gamma_p =2.7$ (solid line) and  $\gamma_p =3.2$
 (dashed line).}
\label{fig:z-regen}       
\end{figure}

Now, making use of Eq.\ (\ref{eq:Inu-prompt-tot}) separately 
below and above the CR spectral breaks, inserting the nucleon regeneration moments
$Z_{p{\rm -air}}^N$ calculated with the QGSJET-III model and shown in  
Fig.\  \ref{fig:z-regen},
and using charmed hadron decay moments $Z^{\rm dec}_{h_c\rightarrow \nu_{\mu}}$
complied in \cite{tig96},
we obtain prompt muon neutrino flux in the 
10--100 TeV and 1--10 PeV
energy intervals, 
        \begin{figure*}[t]
\centering
\includegraphics[height=6cm,width=0.8\textwidth]{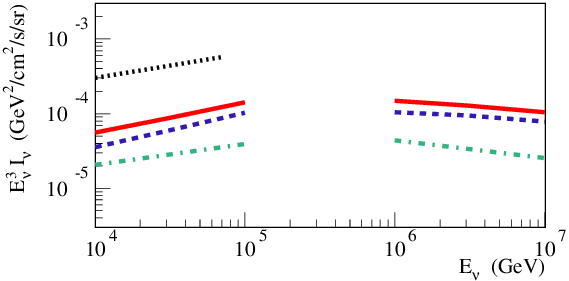}
\caption{Calculated prompt atmospheric muon neutrino flux in the 
10--100 TeV and 1--10 PeV energy intervals (red solid lines);
 partial contributions of the perturbative and nonperturbative charm
production are plotted as blue dashed and green dash-dotted lines, respectively. The IceCube upper limit on the prompt atmospheric $\nu_{\mu}$ background 
\cite{aar16} is shown by black dotted line.}
\label{fig:nu-flux}       
\end{figure*}%
plotted in Fig.\  \ref{fig:nu-flux}, with the 
 partial contributions of the perturbative charm production and of the
 intrinsic charm being shown as well. While
the  perturbative charm production  dominates the  prompt $\nu_{\mu}$ flux both at low and high
 energies,   the neutrino
yield from intrinsic charm is also very significant,
for the chosen model parameters,
 amounting  to $\sim 25-35$\% of the total flux, depending on the energy range.

We can also compare the calculated muon neutrino flux to the IceCube
 upper limit on the prompt
 $\nu_{\mu}$ background \cite{aar16}, shown in  Fig.\  \ref{fig:nu-flux} by the dotted line. As one can see
 in the Figure, the total predicted  prompt muon neutrino flux
  is substantially below this limit.
 However, with a further increase of the statistics    of the measurements and a refinement
 of the corresponding analysis methods (see, e.g., \cite{boe23}),
  a direct detection of the prompt atmospheric neutrino flux looks quite feasible.
 Another point to keep in mind is the potential of IceCube to constrain the IC content of the
 proton, envisaged in \cite{lah17}. With the contribution to the atmospheric neutrino
 fluxes from the perturbative charm   production, being seriously constrained by LHCb data,
 even the current IceCube upper limit on the prompt
 $\nu_{\mu}$ background does not leave too much space for the IC part,
 while much tighter constraints
 may be expected in the future.

     A  natural question concerns the uncertainties regarding our results for the prompt
 atmospheric neutrino background. While this can not be answered quantitatively without a dedicated
 analysis which goes beyond the scope of the current work, we can discuss here the main sources of
 such uncertainties. 
 
 Generally   predictions both for the conventional and for 
 the prompt atmospheric
  neutrino backgrounds depend substantially on the parametrizations of the primary cosmic ray fluxes,
  employed in the corresponding calculations. 
  This is not surprising since such 
  parametrizations typically relied at very high energies ($E> 1$ PeV) on
   experimental results for the total cosmic
  ray flux, supplementing those by some plausible assumptions regarding the composition and the nature of potential sources of
  high energy cosmic rays. Consequently, the corresponding predictions for the primary proton flux
  at such high energies appeared to be rather uncertain and varied considerably from one parametrization
  to another. Since primary protons make $\sim 70$\% of the prompt atmospheric neutrino background [cf.\
  Eq.\ (\ref{eq:Inu-prompt-tot})], we prefered to rely  on  the data of the  KASCADE-Grande \cite{ape13b} 
  and Ice-Cube \cite{aar19} experiments, regarding the fluxes of CR protons and helium. Yet
  the recently reported rather precise measurement of the primary proton flux by the LHAASO experiment
  \cite{cao25} should provide a much more solid basis for calculations of the atmospheric neutrino background
  and reduce thereby considerably the dependence of the results on the parametrizations of CR fluxes.
  
  Coming now to the impact of uncertinties regarding the description of charmed hadron production, the
  least constrained remains the nonperturbative intrinsic charm contribution. As discussed in Section \ref{res.sec},
  our normalization for  the IC   content  of the proton likely corresponds to the upper bound.
  How much smaller this normalization should be depends on a model treatment of the hadronization process,
  notably, on the strength of the ``color drag'' effect \cite{nor00} in the corresponding string fragmentation
  procedures. The perturbative mechanism typically produces charm (anti)quarks at small (absolute) values
  of the rapidity in the c.m.\ frame, hence, at relatively small $x$. However, when such (anti)quarks
  are connected by strings (tubes) of color field to valence partons of the incident proton, the color
  force may ``drag'' them towards higher $x$. Therefore, if the mechanism is strong enough,
  the relevant experimental data on charm production at fixed target energies can be more or less 
  described by the perturbative mechanism alone, thereby leaving little room for the IC contribution.
  Naively, one could have expected that the total, perturbative plus nonperturbative,
  charm production changes little in such a case. This is, however, not true because of the different
  energy dependence of the two contributions (cf.\ Fig.\ \ref{fig:sig-cc}). Consequently, a model
  characterized by a stronger ``color drag'' effect would predict a higher prompt neutrino background
  (see, e.g., the corresponding discussion in \cite{bha16}).
  Added to these are uncertainties regarding the large $x$ behaviour of the gluon PDF of the proton.
  While this PDF is now substantially constrained in the low $x$ limit by the LHCb measurements 
  \cite{aai13,aai16,aai17}, the large $x$ behaviour which impacts strongly the results
   (cf.\  Fig.\ 3 in \cite{ogs23})
  remains much less constrained.  
  Taking all this into account, the uncertainty of our calculation
 of the prompt atmospheric neutrino background may amount up to a factor of a few.

\section{Discussion and conclusions\label{out.sec}}

We presented a description of charm production in hadronic collisions, in the framework of the 
QGSJET-III MC generator, taking into consideration both the perturbative generation of charm
and the nonperturbative intrinsic charm contribution. The former is treated using the LO pQCD formalism,
with the magnitude of the contribution being controlled by a $K$-factor, $K_{\rm f}^c$, whose value 
has been fixed based on a comparison with LHCb data on $D$ meson production. For IC, we employ
a Regge-based formalism and, considering a Good-Walker expansion of  hadron wave functions, assume
that   IC   content of a particular GW Fock state is inverse proportional to the transverse
size squared of that state. 
The relative importance of intrinsic charm is governed by a parameter specifying the IC content
of the largest GW state considered
 of the proton, whose value has been fixed based on a comparison with 
experimental data on $\Lambda_c$ baryon and $D$ meson
 production at large values of Feynman $x$.

The extremely low IC   content of the proton, smaller than one per mille, obtained by us, may
seem surprising, in particular, in view of previous theoretical studies involving both an analysis 
of proton structure functions (SFs) (e.g., \cite{hob14,jim15}) or based on 
comparisons with charmed hadron production data
(e.g., \cite{gia18,mac20}), which typically yielded an order of magnitude higher values. Regarding
a SF analysis, the main support for a significant IC content of the proton comes from EMC studies of  muon interactions in an iron target \cite{aub80,aub82,aub83}. 
However, there was a serious caveat in the corresponding interpretations,
revealed in \cite{blu16}. To resolve very compact IC states, the virtual photon ``probe'' has to
be pointlike enough, i.e., to be characterized by a sufficiently high virtuality, 
$q^2\gtrsim 150$ GeV$^2$, steeply rising with an increase of Bjorken $x$,
which exceeds the values corresponding to
 the EMC measurements, in the kinematic range covered
by the   experiment \cite{blu16}. Similar concerns apply to
theoretical   studies of Refs.\  \cite{gia18,mac20}, where one considered a direct hard scattering of IC
(anti)quark off a gluon or sea quark coming from the partner proton (nucleus): one allowed
the momentum transfer squared for such a scattering to be as low as few GeV$^2$, 
which gave rise to an
``explosion'' of the corresponding cross section in the small $p_{\perp}$ limit. Taking
properly into account the above-mentioned kinematic 
 condition for resolving the intrinsic charm,
the momentum transfer for such a hard scattering should be quite high, thereby making the
direct scattering contribution negligibly small. In contrast, we followed the approach
of Ref.\ \cite{bro89,bro92}, assuming that  a scattering of {\em any} parton constituent of
incident hadron is sufficient to destroy the coherence of the initial parton configuration of
(a Fock state of) the hadron and thereby ``to free the intrinsic charm from its
 virtual state'' \cite{bro92}.
In such a case, the contribution of IC to charmed hadron production is approximately
proportional to the total inelastic cross section for a reaction.

Regarding the importance of charmed hadron production for calculations of   atmospheric
neutrino background, the corresponding perturbative contribution can be conveniently studied
at the level of CR spectrum-weighted moments  for $c$($\bar c$) production \cite{ogs23}.
Comparing the energy dependence of these moments between our LO treatment and NLO calculations,
the obtained differences appeared to be comparable to theoretical uncertainties of the latter,
obtained by varying the factorization and renormalization scales.

However, for combing both the pertirbative and nonperturbative contributions, one can only factorize
out the decay moments for charmed hadrons, hence, to deal with production moments of those
various hadrons: because the hadronization of IC (anti)quarks generally proceeds differently
than for perturbatively generated ones. Comparing the energy-dependence of such moments, we
observed, as expected, a rather flat one for the IC contribution, while having a pronounced 
energy-rise of the moments corresponding to perturbative generation of charm,   driven
by the low $x$ increase of the gluon PDF of the proton. Consequently,
as noticed earlier in  \cite{ogs23}, the perturbative contribution to the prompt atmospheric
neutrino flux is characterized by a steeper energy-rise than the nonperturbative one.


To conclude, we developed a MC treatment of charmed hadron production, comprising the corresponding
perturbative and nonperturbative mechanisms.
Since the magnitude of the two contributions is
controlled by independent adjustable parameters, one can use this fact for studying the 
uncertainties of the corresponding predictions and for further retuning the model, based on 
new experimental data both from accelerators and from astrophysical observatories.

\subsection*{Acknowledgments}

The authors acknowledge useful discussions with M.\ V.\ Garzelli and other participants of the
2025 MIAPbP program ``Event Generators at Colliders and Beyond Colliders''.
S.O.\ acknowledges  financial support from the Deutsche Forschungsgemeinschaft 
(projects 465275045 and 550225003).  G.S.\ acknowledges
support by the Bundesministerium f\"ur Bildung
und Forschung, under grants 05A20GU2 and 05A23GU3, and
by the Deutsche Forschungsgemeinschaft  under    
  Germany's Excellence Strategy -- EXC 2121 ``Quantum Universe'' -- 390833306.


\begin{thebibliography}{99}

\bibitem{gai95} 
T.\ K.\ Gaisser, F.\ Halzen, and T.\ Stanev,
 {\em Particle astrophysics with high-energy neutrinos},
 Phys.~Rep.\  {\bf   258},   173 (1995).

\bibitem{nas89} P.\ Nason, S.\ Dawson, and R.\ K.\ Ellis,
  {\em The one particle inclusive differential cross section for heavy
  quark production in hadronic collisions},
Nucl.\ Phys.\ B {\bf 327},  49 (1989).

\bibitem{bee91} W.\ Beenakker, W.\ L.\ van Neerven, R.~Meng, G.\ A.\ Schuler, and J.\ Smith,
  {\em QCD corrections to heavy quark production in hadron hadron collisions},
Nucl.\ Phys.\ B {\bf 351},  507 (1991).

\bibitem{aai13}  R.\ Aaij  et al.\ (LHCb Collaboration),
  {\em Prompt charm production in $pp$ collisions at $\sqrt{s}=7$ TeV},
Nucl.\ Phys.\ B  {\bf 871}, 1 (2013).

\bibitem{aai16}  R.\ Aaij  et al.\ (LHCb Collaboration),
  {\em Measurements of prompt charm production cross-sections
   in  $pp$  collisions at  $\sqrt{s}=13$ TeV},
 JHEP {\bf 03},  159 (2016).

\bibitem{aai17}  R.\ Aaij  et al.\ (LHCb Collaboration),
  {\em Measurements of prompt charm production cross-sections
   in  $pp$  collisions at  $\sqrt{s}=5$ TeV},
 JHEP {\bf 06},  147 (2017).
 
\bibitem{gau15} R.\ Gauld, J.\ Rojo, L.\ Rottoli, and J.\ Talbert,
   {\em Charm production in the forward region: constraints on the small-$x$
   gluon and backgrounds for neutrino astronomy},
 JHEP {\bf 11}, 009 (2015).

 \bibitem{zen15} O.\  Zenaiev et al.\ (PROSA Collaboration),
{\em Impact of heavy-flavour production cross sections measured 
by the LHCb experiment on parton distribution functions at low x},
 Eur.\ Phys.\  J.\ C {\bf 75}, 396 (2015).

 \bibitem{gau17} R.\ Gauld and J.\ Rojo, 
   {\em Precision determination of the small-$x$ gluon from charm production at LHCb},
 Phys.\  Rev.\ Lett.\ {\bf 118}, 072001 (2017).

 \bibitem{zen20}  O.\ Zenaiev,  M.\ V.\ Garzelli, K.\ Lipka, S.-O.\ Moch, 
A.\ Cooper-Sarkar, F.\ Olness, A.\ Geiser,  and G.\ Sigl (PROSA Collaboration), 
 {\em  Improved constraints
on parton distributions using LHCb, ALICE and HERA heavy-flavour measurements
and implications for the predictions for prompt atmospheric-neutrino fluxes},
 JHEP {\bf 04}, 118 (2020).

 \bibitem{nel13} R.\ E.\ Nelson, R.\ Vogt, and A.\ D.\ Frawley,
  {\em Narrowing the uncertainty on the total charm cross section and its effect 
  on the $J/\psi$ cross section},
 Phys.\  Rev.\ C {\bf 87},  014908 (2013).  

\bibitem{bhps80} S.\ J.\ Brodsky, P.\ Hoyer, C.\ Peterson, and N.~Sakai,
  {\em The intrinsic charm of the proton},
 Phys.\  Lett.\ B {\bf 93}, 451 (1980).
 
\bibitem{bps81} S.\ J.\ Brodsky,  C.\ Peterson, and N.\ Sakai, {\em Intrinsic
heavy-quark states},
 Phys.\  Rev.\ D {\bf 23}, 2745 (1981).

\bibitem{hob14} T.\ J.\ Hobbs, J.\ T.\ Londergan, and W.\ Melnitchouk,
{\em Phenomenology of nonperturbative charm in the nucleon},
 Phys.\  Rev.\ D {\bf 89}, 074008 (2014).

\bibitem{jim15} 
  P.\ Jimenez-Delgado, T.\ J.\ Hobbs, J.\ T.\ Londergan, and W.\ Melnitchouk,
 {\em  New limits on intrinsic charm in the nucleon from global analysis of parton distributions},
   Phys.\ Rev.\ Lett.~{\bf  114}, 082002  (2015).	

\bibitem{bed19}  V.\ A.\ Bednyakov,  S.\ J.\ Brodsky,  A.~V.~Lipatov,  G.\ I.\ Lykasov, 
 M.\ A.\ Malyshev,  J.~Smiesko, and S.\ Tokar,
  {\em  Constraints on the intrinsic charm content of the proton from recent ATLAS data},
 Eur.\ Phys.\  J.\ C {\bf 79}, 92 (2019).
  
 \bibitem{ost24a} S.\ Ostapchenko, 
 {\em QGSJET-III model of high energy hadronic interactions: The formalism},
 Phys.\  Rev.\ D {\bf  109},   034002 (2024).  
 
   \bibitem{ost24b} S.\ Ostapchenko, 
 {\em QGSJET-III model of high energy hadronic interactions: II. Particle
production and extensive air shower characteristics},
 Phys.\  Rev.\ D {\bf  109},   094019  (2024). 
  
   \bibitem{bar91} G.\ Bari  et al.,
 {\em A Measurement of $\Lambda_c^+$ baryon production in proton proton 
 interactions at s**(1/2) = 62 GeV},
 Nuovo Cim.\ A {\bf  104},    571 (1991). 

\bibitem{aar16}  M.\ G.\ Aartsen et al.\ (IceCube Collaboration), 
 {\em 	 Observation and characterization of a cosmic muon neutrino flux
from the northern hemisphere using six years of IceCube data}, 
Astrophys.\ J.\ {\bf 833}, 3  (2016).

\bibitem{gri68}
 V.~N.~Gribov,  
{\em   A reggeon diagram technique}, 
 Sov.~Phys.~JETP {\bf 26}, 414 (1968). 
 
\bibitem{gri69}
 V.\ N.\ Gribov, 
 {\em Glauber corrections and the interaction between high-energy 
hadrons and nuclei},
  Sov.\ Phys.\ JETP  {\bf 29}, 483, (1969). 
  
	
\bibitem{gri72}
 V.\ N.\ Gribov  and L.\ N.\ Lipatov, 
 {\em Deep inelastic e p scattering in perturbation theory},
 Sov.\ J.\ Nucl.\ Phys.\   {\bf   15}, 438 (1972).

\bibitem{alt77}
 G.\  Altarelli and  G.\  Parisi,
 {\em Asymptotic freedom in parton language}, 
 Nucl.\ Phys.\ B   {\bf  126}, 298 (1977).

\bibitem{dok77}
 Yu.\ L.\ Dokshitzer,
 {\em Calculation of the structure functions for deep inelastic 
scattering and e+ e- annihilation by perturbation theory in 
quantum chromodynamics}, 
 Sov.\ Phys.\ JETP   {\bf  46}, 641 (1977).

\bibitem{dre99}
 H.\ J.\ Drescher, M.\ Hladik,  S.\  Ostapchenko,  and K.\  Werner,  
{\em A Unified treatment of high-energy interactions}, 
 J.\ Phys.\ G  {\bf  25},   L91 (1999).
 
\bibitem{dre01}   	
 H.\ J.\ Drescher, M.\ Hladik,  S.\  Ostapchenko, T.\ Pierog,  and K.\  Werner,  
{\em Parton based Gribov-Regge theory},
 Phys.~Rep.\  {\bf  350},    93 (2001).
 
\bibitem{ost02}
 S.\  Ostapchenko,  H.\ J.\ Drescher,  F.\ M.\  Liu, T.\ Pierog,  and K.\  Werner,  
{\em Consistent treatment of soft and hard processes in hadronic interactions},
  J.\ Phys.\ G  {\bf 28}, 2597 (2002).

\bibitem{ost06} 
S.~Ostapchenko,
  {\em On the re-summation of enhanced Pomeron diagrams},
 Phys.\ Lett.\ B  {\bf 636},   40 (2006).

\bibitem{ost08} S.\ Ostapchenko,
 {\em Enhanced Pomeron diagrams: Re-summation of unitarity cuts},
  Phys.~Rev.~D     {\bf 77}, 034009 (2008).

\bibitem{ost10} S.\ Ostapchenko,
 {\em Total and diffractive cross sections in enhanced Pomeron scheme},
 Phys.~Rev.~D    {\bf  81}, 114028 (2010). 

     \bibitem{agk74}
 V.\ A.\ Abramovsky,  V.\ N.\ Gribov, and  O.~V.~Kancheli, 
 {\em Character of inclusive spectra and fluctuations produced in
 inelastic processes by multi-pomeron exchange}, 
  Sov.~J.~Nucl.~Phys.\   {\bf   18}, 308 (1974).
	
\bibitem{ost11} S. Ostapchenko, 
 {\em Monte Carlo treatment of hadronic 
interactions in enhanced Pomeron scheme: QGSJET-II model}, 
 Phys.\  Rev.\ D {\bf 83},  014018 (2011).
   
\bibitem{col89}
J.\ C.\ Collins,  D.\ E.\ Soper,  and G.\ F.\ Sterman,
 {\em Factorization of hard processes in QCD},
Adv.\ Ser.\ Direct.\ High Energy Phys.\  {\bf 5}, 1 (1989).

\bibitem{ost06a}
 S.~Ostapchenko,   
 {\em  Nonlinear screening effects in high energy hadronic interactions},
    Phys.~Rev.~D~{\bf 74}, 014026 (2006).

  \bibitem{ost16}
S.~Ostapchenko and   M.~Bleicher, 
 {\em Double parton scattering: impact of nonperturbative 
parton correlations},
 Phys.\ Rev.\  D  {\bf 93},  034015 (2016).

\bibitem{ell96}
R.~K.~Ellis, W.~J.~Stirling and B.~R.~Webber,
 {\em QCD and collider physics},
Camb.\ Monogr.\ Part.\ Phys.\ Nucl.\ Phys.\ Cosmol.~{\bf 78} (1996), 1,
Cambridge University Press, 2011.

\bibitem{kai03} A.\ B.\ Kaidalov, {\em $J/\psi$ $c\bar c$ production in $e^+e^-$
and hadronic interactions},
JETP Lett.\ {\bf 77}, 349 (2003).

\bibitem{kai86} A.\ B.\ Kaidalov and O.\ I.\ Piskunova, 
{\em Inclusive spectra of baryons in the Quark-Gluon String model},
Z.\ Phys.\  C {\bf 30}, 145 (1986).

\bibitem{kai87}
A.~B.~Kaidalov,  
{\em Quark and diquark fragmentation functions in the model of quark gluon strings},
  Sov.~J.~Nucl.~Phys.~{\bf 45}, 902 (1987).
 
\bibitem{bro89} S.\ J.\ Brodsky and  P.\ Hoyer,
  {\em Nucleus as a color filter in QCD: Hadron production in nuclei},
 Phys.\  Rev.\ Lett.\ {\bf 63}, 1566 (1989).
 
\bibitem{bro92} S.\ J.\ Brodsky,  P.\ Hoyer, A.\ H.\ Mueller, and W.-K.\ Tang,
  {\em New QCD production mechanisms for hard processes at large $x$},
Nucl.\ Phys.\  B {\bf 369},  519 (1992).

\bibitem{ogs23}  S.\ Ostapchenko, M.\ V.\ Garzelli, and G.\ Sigl,
 {\em On the prompt contribution to the atmospheric neutrino flux},
 Phys.\  Rev.\ D {\bf 107},  023014 (2023).

\bibitem{goo60}
M.\ L.\ Good and W.\ D.\ Walker,
  {\em Diffraction disssociation of beam particles},
Phys.\ Rev.\ {\bf 120}, 1857 (1960).

\bibitem{fra08}
L.~Frankfurt, M.~Strikman, D.~Treleani, and C.~Weiss,
   {\em Evidence for color fluctuations in the nucleon in high-energy scattering},
    Phys.\ Rev.\ Lett.~{\bf 101}, 202003 (2008).	
 
\bibitem{cha87} P.\ Chauvat et al.\ (R608 Collaboration),
 {\em Production of $\Lambda_c$ with large $x_{\rm F}$ at the ISR},
  Phys.\ Lett.\ B  {\bf  199}, 304 (1987).
  
\bibitem{gar02}  F.\ G.\ Garcia et al.\ (SELEX Collaboration),
{\em Hadronic production of $\Lambda _c$ from 600 GeV/c $\pi ^-$, $\Sigma ^-$
and $p$ beams},
 Phys.\  Lett.\ {\bf B528}, 49 (2002).
 
 \bibitem{bha16} A.\ Bhattacharya, R.\ Enberg, M.\ H.\ Reno,  I.\ Sarcevic,
 and A.\ Stasto,  {\em Prompt atmospheric neutrino fluxes: perturbative 
QCD models and nuclear effects},
 JHEP {\bf 11}, 167 (2016).
  
\bibitem{amm88} R.\ Ammar et al.\  (LEBC-MPS Collaboration) 
 {\em  D-Meson production in 800-GeV/c $pp$ interactions},
 Phys.\  Rev.\ Lett.\ {\bf 61},  2185 (1988).

\bibitem{ct14} S.\ Dulat, T.-J.\ Hou, J.\ Gao, M.\ Guzzi,  J.\ Huston,  
P.\  Nadolsky, J.\ Pumplin,  C.\ Schmidt, D.\ Stump, and C.-P.\ Yuan,
  {\em New parton distribution functions from a global analysis of quantum 
chromodynamics},
 Phys.\  Rev.\ D {\bf 93}, 033006 (2016).

\bibitem{bal11}  R.\ D.\ Ball et al.\ (NNPDF Collaboration),
  {\em Parton distributions from high-precision collider data},
  Eur.\ Phys.\  J.\ C {\bf 77} (2017) 663.

  
\bibitem{kul59} G.\ Kulikov and G.\ Khristiansen, {\em On the size spectrum
of extensive air showers},
Sov.\ Phys.\ JETP  {\bf 8}, 441 (1959).

\bibitem{ant05}  T.\ Antoni et al.\ (KASCADE Collaboration), {\em KASCADE
measurements of energy spectra for elemental groups of cosmic rays: 
Results and open problems},
Astropart.\ Phys.\   {\bf 24}, 1 (2005).

\bibitem{ape11}   W.\ D.\ Apel et al.\ (KASCADE-Grande Collaboration), 
{\em Knee-like structure in the spectrum of the heavy component of 
cosmic rays observed with KASCADE-Grande},
 Phys.\  Rev.\ Lett.\ {\bf 107}, 171104 (2011).

\bibitem{ape13b}  W.\ D.\ Apel et al.\ (KASCADE-Grande Collaboration), 
{\em KASCADE-Grande
measurements of energy spectra for elemental groups of cosmic rays},
 Astropart.\ Phys.\   {\bf 47}, 54 (2013).

\bibitem{ape13a}  W.\ D.\ Apel et al.\ (KASCADE-Grande Collaboration), 
{\em Ankle-like feature in the energy spectrum of light elements of 
cosmic rays observed with KASCADE-Grande},
 Phys.\  Rev.\ D {\bf 87}, 081101 (2013).

\bibitem{aar19}  M.\ G.\ Aartsen et al.\ (IceCube  Collaboration), {\em Cosmic ray
spectrum and composition from PeV to EeV using 3 years of data from
IceTop and  IceCube}, Phys.\ Rev.\  D {\bf 100}, 082002 (2019).


\bibitem{tig96} M.\ Thunman, G.\ Ingelman, and P.\ Gondolo, {\em Charm
production and high energy atmospheric muon and neutrino fluxes},
Astropart.\ Phys.\   {\bf 5}, 309 (1996).

\bibitem{boe23}  J.\ Boettcher et al.\ (IceCube  Collaboration), 
{\em Search for the Prompt Atmospheric Neutrino Flux in IceCube},
  PoS  {\bf ICRC2023}, 1068 (2023) [arXiv:2309.07560 [astro-ph.HE]].

\bibitem{lah17} R.\ Laha and S.\ J.\ Brodsky, 
  {\em IceCube can constrain the intrinsic charm of the proton},
 Phys.\  Rev.\ D {\bf 96}, 123002 (2017).
  
\bibitem{cao25}  Z.\ Cao et al.\ (LHAASO Collaboration), 
{\em   First Identification and Precise Spectral Measurement of the 
Proton Component in the Cosmic-Ray `Knee'},
 [arXiv:2505.14447 [astro-ph.HE]].
         
\bibitem{nor00} 
 E.\ Norrbin and T.\ Sjostrand,
    {\em  Production and hadronization of heavy quarks},
   Eur.\ Phys.\  J.\ C {\bf 17}, 137 (2000).


\bibitem{gia18} A.\ V.\ Giannini, V.\ P.\ Goncalves,  and F.~S.~Navarra,
 {\em Intrinsic charm contribution to the prompt 
atmospheric  neutrino flux},
 Phys.\  Rev.\ D {\bf 98}, 014012 (2018).

\bibitem{mac20} R.\ Maciu\l{}a and A.\ Szczurek, 
  {\em  Intrinsic charm in the nucleon and charm production at large
  rapidities in collinear, hybrid and $k_T$-factorization approaches},
 JHEP {\bf 10}, 135 (2020).
 
\bibitem{aub80}   J.\ J.\ Aubert et al.\ (EMC Collaboration),
{\em   A study of dimuon events in 280 GeV muon interactions},
Phys.\ Lett.\ B  {\bf  94},  96  (1980).

\bibitem{aub82}   J.\ J.\ Aubert et al.\ (EMC Collaboration),
{\em   An experimental limit on the intrinsic charm component of the nucleon},
Phys.\ Lett.\ B  {\bf  110},  73  (1982).
    
\bibitem{aub83}   J.\ J.\ Aubert et al.\ (EMC Collaboration),
{\em  Production of charmed particles in 250 GeV $\mu^+$-iron interactions},
    Nucl.\ Phys.\ B  {\bf  213}, 31 (1983).
    
\bibitem{blu16}
J.\ Bl\"umlein,
 {\em A Kinematic condition on intrinsic charm},
 Phys.\ Lett.\ B  {\bf  753},  619  (2016).

\end{thebibliography}
\end{document}